# The Dalton Minimum and John Dalton's Auroral Observations

Sam M. Silverman

18 Ingleside Road

Lexington, MA

silvermansam111@gmail.com

Hisashi Hayakawa

(1) Institute for Space-Earth Environmental Research

Nagoya University, Nagoya, 4648601, Japan

(2) Institute for Advanced Researches

Nagoya University, Nagoya, 4648601, Japan

(3) UK Solar System Data Centre, Space Physics and Operations Division, RAL Space, Science and Technology Facilities Council, Rutherford Appleton Laboratory,

Harwell Oxford, Didcot, Oxfordshire, OX11 0QX, UK

hisashi@nagoya-u.jp

Abstract

In addition to the regular Schwabe cycles of approximately 11 y, "prolonged solar activity minima" have been identified through the direct observation of sunspots and aurorae, as well as proxy data of cosmogenic isotopes. Some of these minima have been regarded as grand solar minima, which are arguably associated with the special state of the solar dynamo and have attracted significant scientific interest. In this paper, we review how these prolonged solar activity minima have been identified. In particular, we focus on the Dalton Minimum, which is named after John Dalton. We review Dalton's scientific achievements, particularly in geophysics. Special emphasis is placed on his lifelong observations of auroral displays over approximately five decades in Great Britain. Dalton's observations for the auroral frequency allowed him to notice the scarcity of auroral displays in the early 19th century. We analyze temporal variations in the annual frequency of such displays from a modern perspective. The contemporary geomagnetic positions of Dalton's observational site make his dataset extremely valuable because his site is located in the sub-auroral zone and is relatively sensitive to minor enhancements in solar eruptions and solar wind streams. His data indicate clear solar cycles in the early 19th century and their significant depression from 1798 to 1824. Additionally, his data reveal a significant spike in auroral frequency in 1797, which chronologically coincides with the "lost cycle" that is believed to have occurred at the end of Solar Cycle 4. Therefore, John Dalton's achievements can still benefit modern science and help us improve our understanding of the Dalton Minimum.

## 1. Introduction

Mid-latitude aurorae visually manifest solar eruptions and high-speed solar wind streams interacting with terrestrial magnetic fields (Gonzalez *et al*., 1994; Daglis *et al*., 1999). Their visibility has been intermittently recorded for millennia in various historical documents (Silverman, 1998, 2006; Stephenson *et al*., 2004; Vaquero and Vázquez, 2009). These records provide valuable scientific insights into long-term solar variability and solar eruptions prior to telescopic observations (Silverman, 1992, 2006; Willis and Davis, 2015; Lockwood and Barnard, 2015; Hayakawa *et al*., 2017, 2019a, 2019b), particularly when considered in combination with cosmogenic isotope data (Usoskin *et al*., 2007; Inceoglu *et al*., 2015; Usoskin, 2017; Barnard *et al*., 2018; Miyake *et al*., 2019).

Since the seminal work of de Mairan (1733) on auroral frequency, it has been recognized that there are intervals with considerably diminished auroral activity. In particular, Eddy (1976) has marshaled the evidence for a prolonged solar activity minimum called the Maunder Minimum from 1645 to 1715. This grand minimum has served as a benchmark for long-term solar variability with a significant reduction in sunspot occurrence (Usoskin *et al*., 2015; Vaquero *et al*., 2015; Muñoz-Jaramillo and Vaquero, 2019) and apparent loss of solar coronal streamers (Riley *et al*., 2015; Hayakawa *et al*., 2020c). A significant decrease in auroral visibility in the European sector was significant given the high magnetic latitude of this sector during this time period, which would generally cause one to expect more frequent auroral sightings (*e.g.*, Usoskin *et al.*, 2015; Hayakawa *et al.*, 2020d).

Additional evidence can be found in New England, where an aurora was first observed in December of 1719. Lovering (1867, p. 102) noted that it was unlikely that any conspicuous aurora prior to that date would not have been remarked upon because "The people of New England were too much inclined to exaggerate every unusual phenomenon in the heavens to have overlooked or been silent in regard to a spectacle so strange as the aurora, had they had the opportunity of beholding one." In fact, the New England appearance in 1719 filled the country with alarm and many believed that it was a sign of the second coming of Jesus Christ, and that the final judgment was about to commence. In England, Halley (1716), who observed the aurora of 1716 at the age of 60 y, noted that an aurora had not appeared in that part of England since he was born and that he had expected to die without seeing one. He cited a number of observations regarding aurorae appearing between 1560 and 1581, indicating that it was fairly common during that period (see also Usoskin *et al.*, 2015). Similarly, Celsius in Sweden stated that aurorae had been rarely seen there before 1716, but there were 316 observations between 1716 and 1732. De Mairan (1733) noted a comment by Anderson, writing about Iceland, that the older inhabitants were astonished at the frequent appearance of aurorae compared to former times. In 1737, Zanotti commented that aurorae, formerly rare and almost unknown in Italy, had become very frequent. Halley, Leibnitz, Kirch, Fontenelle, and Maraldi all described the aurorae in the first half of the 18th century as uncommon sights. Cassini, a careful observer, did not note the appearance of any aurorae in the latter half of the 17th century. The great weight of contemporaneous evidence throughout Europe, New England, and what were considered the more northerly latitudes of southern Sweden and Iceland indicates considerably diminished auroral activity during the latter half of the 17th century compared to the first half of the 18th century (Silverman, 1992).

A similar prolonged solar activity minimum from the late 1790s to 1827 has been observed in contemporary sunspot records (Clette *et al.*, 2014; Svalgaard and Schatten, 2016; Muñoz-Jaramillo and Vaquero, 2019; Hayakawa *et al.*, 2020a). This interval also has considerable contemporaneous evidence in the form of auroral data to support this observation. As summarized by Silverman (1992), auroral displays were much less frequently recorded in Great Britain and Ireland, especially during the period from 1810 to 1826, and were also missed from 1807 to 1827 in Paris (France), 1789 to 1801 in Karlsruhe (Germany), and 1797 to 1814 in New England. In fact, this trend has been globally confirmed by the auroral records in the European sector (Krivsky and Pejml, 1988; Schröder *et al.*, 2004; Lockwood and Barnard, 2015) and Canadian observatories (Broughton, 2002), as summarized by Vázquez *et al.* (2016). However, caveats must be considered here because these catalogs contain compiled auroral reports from multiple datasets with different observational practices, resulting in poor homogeneity. Therefore, observational records from individual observers should be considered to understand temporal variability in long-term auroral frequency. Such observers include Thomas Hughes at Stroud (UK) from 1771 to 1813 (Harrison, 2005), Francisco Salvá at Barcelona (Spain) from 1780 to 1825 (Vaquero *et al*., 2010), and Giuseppe Tolado at Padua (Italy) from 1766 to 1797 (Dominguez-Castro *et al*., 2016).

In this context, we examine the auroral observations of John Dalton, after whom the Dalton Minimum was named and whose significant achievements have been emphasized in previous publications (*e.g.*, Silverman, 1992; Lockwood and Barnard, 2015). Because he conducted auroral observations at Kendal up to 1793 and at Manchester after that point (Henry, 1854), his observational records are more sensitive to auroral displays connected to recent geomagnetic storms. Because magnetic storms tend to be less intense under reduced solar activity (*e.g*., Usoskin *et al*., 2015; Vázquez *et al*., 2016), Dalton had a greater opportunity to observe auroral displays over time and has provided us with significantly more data regarding the temporal variability of auroral frequency during the Dalton Minimum compared to the data in other catalogs derived from southern European records (*e.g*., Vaquero *et al*., 2010). Additionally, his observations span almost five decades (1786 to 1834) and chronologically cover a period longer than the entire interval of the Dalton Minimum (1797 to 1827; see Usoskin, 2017). This chronological coverage makes Dalton's records particularly valuable because sunspot observations during the Dalton Minimum are relatively scarce based on gaps between long-term sunspot observers (*e.g*., Svalgaard and Schatten, 2016; Vaquero *et al*., 2016; Hayakawa *et al*., 2020a). Additionally, Dalton's records accommodate the controversial reconstruction of contemporary solar activity (*e.g*., Clette and Lefevre, 2016; Svalgaard and Schatten, 2016; Usoskin *et al*., 2016; Chatzistergos *et al*., 2017) and highlight the possibility of a lost solar cycle around the onset of the Dalton Minimum (*e.g*., Usoskin *et al*., 2009). In this article, we clarify how the Dalton Minimum came to be named after Dalton, highlight Dalton's personal profile and geophysical achievements, and compile and analyze his auroral observations.

An extended review of the auroral background was presented by Siscoe (1980). He expanded auroral databases by incorporating several European and Asian databases, and considered the impact of solar variables, migration of earth's magnetic pole, and variations in the magnetic dipole. Subsequent work has been detailed above.

## 2. Naming of the Grand Minima and the Dalton Minimum

In addition to the regular Schwabe cycles of approximately 11 y, evidence of "prolonged solar activity minima" has been identified in direct sunspot observations, particularly in the period from 1645 to 1715 (*e.g*., Spörer, 1889; Maunder, 1922; see Table 1 and Figure 1). Eddy (1976) named this minimum as the Maunder Minimum after the British astronomer who wrote a seminal paper on the minimum from the middle of the 17th century to the early years of the 18th century (*e.g.*, Maunder, 1922). The Maunder Minimum has been independently confirmed to represent enhancements of $^{14}$C and $^{10}$Be as a result of a weakened solar magnetic field and increased inputs of cosmic rays (Eddy, 1976; see also Usoskin *et al*., 2015).

Table 1: Summary of the names, intervals (onset and end), and references for known prolonged minima. Intervals have been adopted from Usoskin *et al*. (2007) for the three prolonged minima before 1610 and from Usoskin *et al*. (2015) for the Maunder Minimum. References have been abbreviated as YSTCC (the Yosemite Solar Terrestrial Coupling Conference in 1978), E76 (Eddy, 1976), SQ80 (Stuiver and Quay, 1980), and SG80 (Stuiver and Grotes, 1980).

| Name | Onset | End | Origin | Reference |
| --- | --- | --- | --- | --- |
| Dalton Minimum | 1797 | 1827 | John Dalton | YSTCC |
| Maunder Minimum | 1645 | 1715 | Edward Maunder | E76 |
| Spörer Minimum | 1390 | 1550 | Gustav Spörer | E76 |
| Wolf Minimum | 1280 | 1350 | Rudolf Wolf | SQ80 |
| Oort Minimum | 1010 | 1050 | Jan Oort | SG80 |

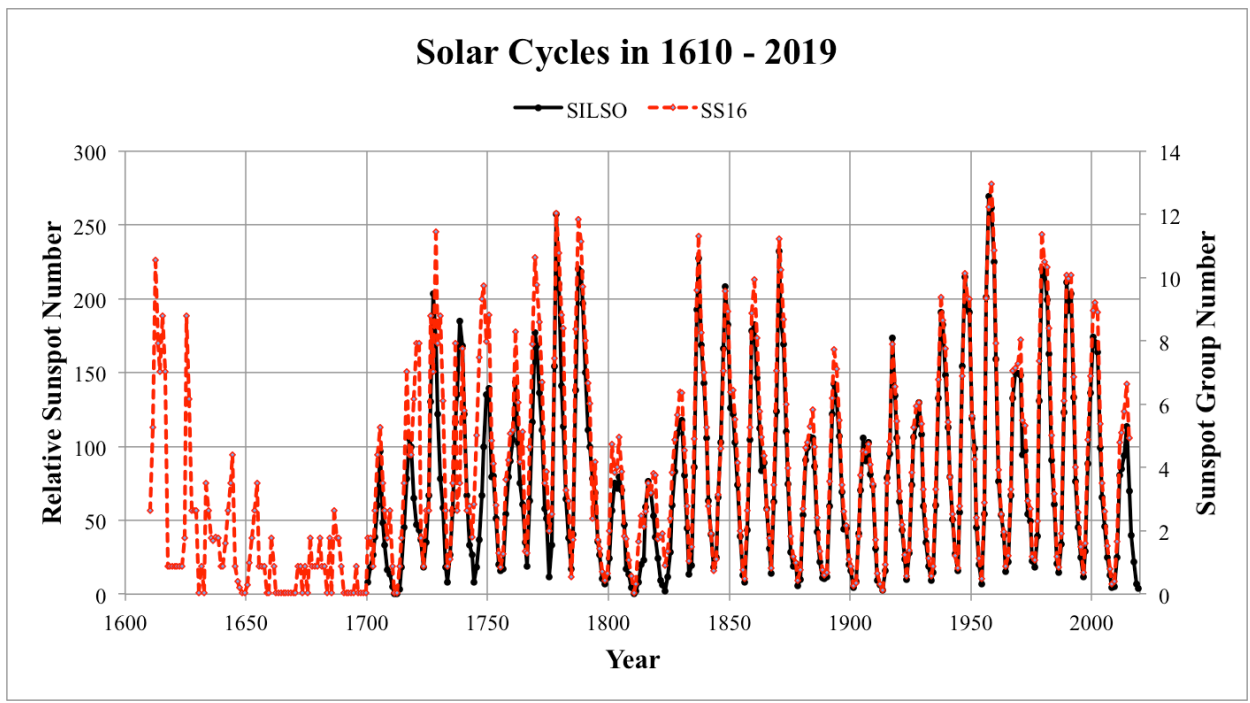

Figure 1. Reconstructions of sunspot group numbers since 1610, visualized with relative sunspot numbers from SILSO (Clette and Lefevre, 2016; black continuous curve) and sunspot group numbers from Svalgaard and Schatten (2016; red broken curve). Two prolonged minima are evident here: the Maunder Minimum (1645 to 1715) and Dalton Minimum (1797 to 1827).

Following the onset of sunspot observations in 1610 (*e.g.*, Clette *et al*., 2014; Arlt and Vaquero, 2020), similar prolonged solar minima have been identified in cosmogenic isotope data based on long-term enhancements of $^{14}$C and $^{10}$Be (*e.g.*, Usoskin, 2017; see Table 1). Major prolonged solar minima over the past millennium gained notoriety in solar physics following a series of studies in the 1970s and 1980s. One of the early minima from 1390 to 1550 was named the Spörer Minimum by Eddy (1976, p. 1196) after Gustav Spörer, who extended studies on the Schwabe cycles and arguably identified a prolonged solar minimum in the late 17th century (Spörer, 1889), but gave its name to Edward Maunder (1851 to 1928). Another prolonged solar activity minimum occurred from 1280 to 1350 and was named the Wolf Minimum by Stuiver and Quay (1980, pp. 16–18), after Rudolf Wolf (1816 to 1893), who established the principles for sunspot numbers and compiled the first comprehensive sunspot database (*e.g.*, Clette *et al*., 2014; Friedli, 2016). A similar prolonged minimum occurred from 1010 to 1050 and was named the Oort Minimum by Stuiver and Grootes (1980, pp. 170–71) after Jan Hendrik Oort (1900 to 1992), who contributed significantly to the fields of galactic studies and radio astronomy. These prolonged solar minima have been elaborately analyzed based on cosmogenic isotopes and are typically interpreted as grand minima in modern scientific studies (*e.g.*, Usoskin *et al*., 2007; Inceoglu *et al*., 2015).

Similar prolonged solar minima have been identified in analyses of the long-term variability of cosmogenic isotopes. These earlier grand minima have been tentatively named as the Greek Minimum (390 to 330 BCE), Homeric Minimum (810 to 720 BCE), Egyptian Minimum (1410 to 1370 BCE), and Sumerian Minimum (3370 to 3300 BCE) based on cosmogenic isotope data (*e.g*., Figure 7 in Landscheidt (1981)). However, these names have not been widely adopted in solar physics and are seldom used in modern scientific discussions.

At the Yosemite Solar Terrestrial Coupling Conference in 1978, John Eddy, George Siscoe and Sam Silverman (the author of the present paper), had an informal meeting to discuss these prolonged solar minima. In the course of this meeting the discussion came around to the various minima of the medieval period and later, focusing largely on the minimum which Eddy had named the Maunder Minimum, after the British astronomer who had written a seminal paper on the minimum from the middle of the seventeenth century to the early years of the eighteenth century (*e.g*., Maunder, 1922). Silverman suggested to Eddy that it would be worthwhile to study the minimum around the beginning of the nineteenth century, since there a greater number of scientific observers, more and better scientific instrumentation, and more scientific institutions involved in such studies, by contrast with the efforts of the previous century. Silverman suggested that Eddy do such a study. He in turn suggested that Silverman do the study. In the course of this discussion the question of what to name this prolonged minimum. Silverman suggested that it be called the Dalton Minimum (see also Hayakawa *et al*., 2020b). His reasons were first, that Dalton had kept a meteorological diary for many years, that this diary include a long series of observations of auroras over many years, and that Dalton was the founder of atomic theory which had become the basis of all later work in what was then called natural philosophy. This suggestion was mutually accepted. Memory is fallible, Silverman here adds one set of facts in support: Eddy was an astronomer, Siscoe was a geophysicist, and Silverman's training and background was as a chemist ending up as a geophysicist, thus Silverman was the one most likely to be familiar with a man best known as a chemist.

However, using Dalton's names was a mutual decision by all three researchers and we believe that the attribution of the name should be given to all three. This is a prime example of how science should work as a collaborative effort. In the next section, we will provide examples of Dalton's work in meteorology and his studies of aurorae, which will demonstrate that he certainly deserves the distinction. The auroral records related to this minimum were subjected to detailed analyses and published by Feynman and Silverman (1980), and Siscoe (1980), as reviewed by Eddy (1988) and Silverman (1992), who explicitly used the term of the Dalton Minimum to describe this solar minimum.

### 3. Dalton and his studies

Dalton's history is best summarized in a contribution he made to a book by a Mr. Roberts called the *Book of Autographs* in the form of a letter on 22 October 1832. He summarized his life as "Attended the village schools, there and in the neighbourhood, till eleven years of age, at which period he had gone through a course of mensuration, surveying, navigation, &c.; began about twelve to teach the village school, and continued it two years; afterwards was occasionally employed in husbandry for a year or more; removed to Kendal at fifteen years of age as assistant in a boarding-school; remained in that capacity for three or four years, then undertook the same school as a principal, and continued it for eight years; whilst at Kendal, employed his leisure in studying Latin, Greek, French, and the mathematics, with natural philosophy ; removed thence to Manchester in 1793 as tutor in mathematics and natural philosophy in the New College; was six years in that engagement, and after was employed as private and public teacher of mathematics and chemistry in Manchester, but occasionally by invitation in London, Edinburgh, Glasgow, Birmingham, and Leeds." (Henry, 1854, p.2).

Dalton was born in Eaglesfield, Cumberland on 6 September 1766 in the north of England (Millington, 1906, p. 5; Smith, 1856). He was educated at a village school until the age of 12, and then immediately began teaching younger students (Millington, 1906, p. 7). He was self-taught thereafter with help from peers such as John Gough, a blind genius. He met Gough after he went to Kendal in 1781. It was at Gough's suggestion that he began to keep a meteorological journal in 1787, which he continued for 57 y, with the last entry being the day before his death in 1844. His first entry into the journal was a description of an aurora. His meteorological observations at Kendal were published in 1793 (Dalton, 1793; second edition 1834). These will be discussed in greater detail below. Smith (1856, p. 10) estimated that Dalton made 40,000 meteorological observations, but also cited an estimate by a Mr. Harland of 200,000 observations (Smith, 1856, p. 278–279). Dalton, having become distrustful of measurements from other individuals, constructed his own instruments, including a barometer, thermometer, and hygroscope (Millington, 1906, p. 12).

The variety of his interests during this period are indicative of his character, having a curiosity for every aspect of nature. Having his curiosity aroused, he would gather facts, either by observations or experiments, and then speculate on possible theories that would provide a description tying the facts together. As a simple example, in a letter to his friend William Alderson, he described the origins of English surnames (Millington, 1906, p. 21–23). He collected botanical specimens and butterflies until 1793 (Angus, 1856, p. 16–17), and also collected insects (Millington, 1906, p. 24). He and his brother offered a course of instruction by subscription. Next, in 1793, he moved to Manchester to teach at the New College. Finding the existing textbooks on grammar inadequate, he wrote his own book on the subject, which was published in 1801.

In Manchester, he became a member of the Manchester Society for Literature and Philosophy. His first paper in their journal, published in 1794, was an article on his color blindness. He subsequently became the Secretary of the Society and then President of the Society, as well as the editor for their journal. As the editor, he would sometimes write essays to fill space when there were insufficient contributions. Much of his original work was published there. Altogether, he published at least 116 papers in the journal (see the listing in Angus, 1856, pp. 253–261). Here, we mention a few of these papers to illustrate the variety of his interests: on the color of the sky: 1795, an essay on the mind: 1798, a paper on grammar: 1800, a paper on winds: 1832, on the specific heat of bodies: 1808, on meteorology: 1812, on the quantity of rain in the previous 25 y: 1818, a paper on indigo: 1823, on 31 y of meteorology in Manchester: 1825, two papers on the height of the aurora on 29 March 1826: 1826, and on an auroral arch on 3 November 1834: 1834. An interesting study in 1788 included "experiments on his own ingesta and egesta, in order to ascertain the weight lost by insensible perspiration: these were published, forty years afterwards, in the memoirs of the Manchester Society, vol. V, p. 303" (Henry, 1854, p.14).

Dalton could often be acerbic in his comments. Commenting on a long paper presented at a meeting after being asked his opinion, he said "well, gentlemen, I daresay this paper is very interesting to those who take an interest in it" (Millington, 1906, p.164). Smith (1856, p. 71) defined Dalton's character as follows: "he was a simple inquirer into nature, his enthusiasm rose only in her presence, his life was devoted to her study."

## 4. Dalton's geophysical studies

Dalton's early geophysical studies were conducted while he resided at Kendal. The corresponding observations and interpretations of results were published in 1793 (Dalton, 1793, 1834), shortly before his move to Manchester. A second edition was published in 1834, which consisted of a verbatim copy of the first edition with an appendix containing some observations on clouds, thunder, and meteors, with special focus on aurorae, including a listing of aurorae over the previous 40 y, which were additions to the first-edition listing of those he had observed in Kendal.

His motivation for this volume was to report on "… several things occurred to me which were new, at least to myself, and which throw light on the different branches of natural philosophy, and of meteor ology in particular," in the form of essays "… in which are also given, such useful discoveries and observations of others as seemed necessary to be known, in order to form a proper idea of the present state of the science, and of the improvements that are yet to be made in it." These observations were made by Dalton at Kendal and by his friend Mr. Crosthwaite at Keswick. The observations of aurorae were new and, in some respects, original. As examples of work included in this volume, there was a second essay on the theory of trade winds, a third essay on variations in barometric pressure, and a seventh essay on the relationship between barometric pressure and rain. An eighth essay was published on the aurora borealis, including a discussion of Halley's work in the early 18th century and the use of a terrella to demonstrate the influence of magnetism on a spherical surface. An appendix used Dalton's barometrical observations to derive the heights of Kendal and Keswick above sea level, as well as the heights of mountains in the area around Keswick.

For the remainder of this paper, we will focus on Dalton's work on aurorae. Aurorae were a major interest of his and approximately one-third of the pages in the 1834 edition focused on aurorae.

## 5. Dalton's catalog and descriptions of aurorae

Dalton recorded a catalog of aurorae observed by himself at Kendal (N54°20, W002°45) and by Crosthwaite at Keswick (N54°36, W003°09) from May of 1786 to May of 1793 (ibid. p. 54). In the second edition, he supplemented this list with a list of aurorae observed by himself at Manchester (N53°29, W002°14) and by others in England from 1793 to 1833 (ibid. p. 217 ff).

Among the detailed descriptions of aurorae, it is noteworthy that for the aurora n March 30, 1793, the estimated height of the lower extremity of the beams was 62 English miles (ibid. p.71). For the same aurora, Dalton noted that the auroral arches were all at right angles to the magnetic meridian (ibid. p. 70). He noted that the aurora on May 24, 1788 was "uncommonly brilliant" (ibid. p. 55). Regarding auroral influence on the magnetic needle, he noted that "I have never observed any considerable fluctuation of the needle in any evening but when there was an aurora visible" (ibid. p. 73).

In late 1834, Dalton noted that "…in fact, the light of the aurora exactly corresponds with that of the electric spark, when sent through a tube in which the air has been rarefied to as high a degree as can be effected by a good air-pump" (ibid. p. 244).

Observers of aurorae from the 18th century to the early part of the 20th century often reported a dark segment below the auroral arches. In a footnote, Dalton observed that (ibid. p. 233) "

This great black cloud which the author describes, was nothing more than the one which is usually seen (or rather imagined) under the low brilliant auroræ, and through which, stars of the first magnitude may be seen. It is a mere contrast of light and shade, and the cloud vanishes when the aurora does" (ibid. p. 233).

## 6. Frequency of aurorae

Dalton noted that aurora "has appeared frequently to all the northern parts of Europe since the year 1716, though it seems to have been a rare phenomenon before that time" (ibid. p.53). This phenomenon is now recognized as the Maunder Minimum. He notes the scarcity of aurorae in the early 19th century, which is currently in his honor as the Dalton Minimum. He observed that there were no aurorae in 1798, 1807, 1809 to 1813, or 1822 to 1824. (ibid. p. 217). In a note, he added that in 1834, "…the late period being one in which the phenomena were less common than they had been previously." He also noted that the aurorae were subject to long periods of intermission, stating (ibid. p. 227) "…the phenomena are subject to long periods of intermission. In some periods, they are rarely seen for great parts of an age; in other periods, they may be viewed 30, 40, or 50 times in each successive year, as was the case forty years since in Great Britain" (ibid. p. 227). Based on the auroral catalogs in Dalton's book (1834), Figure 2 clearly presents the minimum auroral frequency at the end of the 18th and beginning of the 19th centuries.

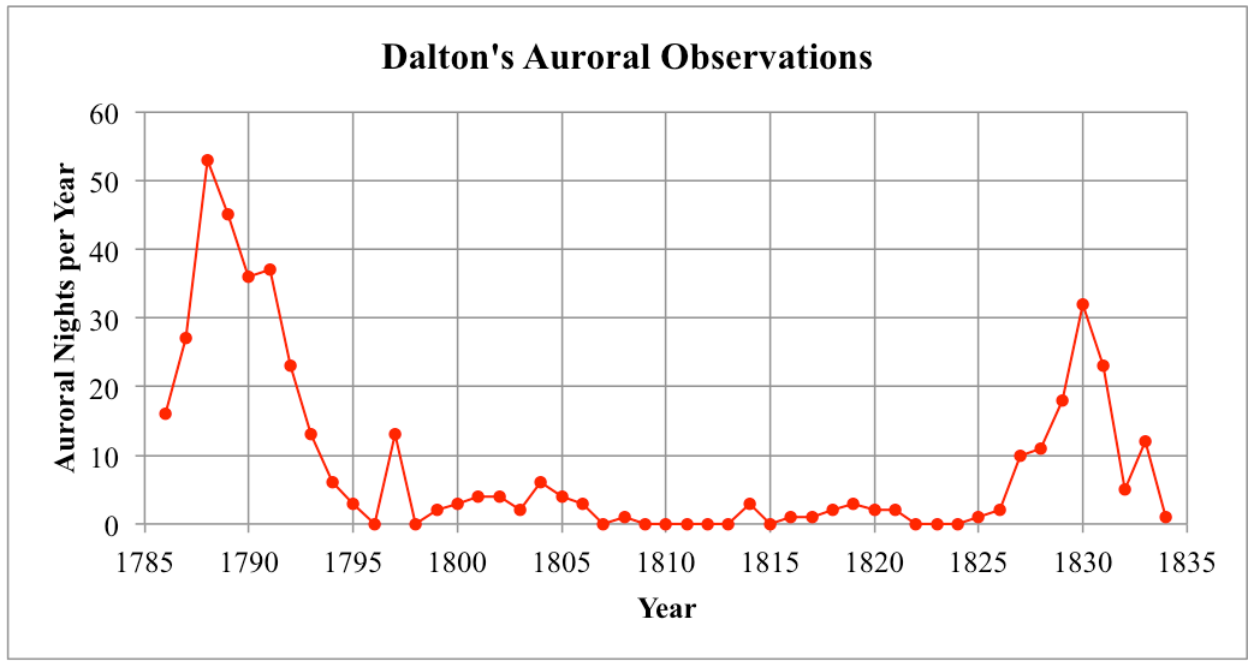

Figure 2. Frequency of aurorae in Great Britain and Ireland according to Dalton's catalogs (1834). A prolonged minimum is clearly evident.

As possible reasons for these variations and in consideration for the strong magnetic connection with aurorae, he suggested that "… a very plausible reason for the appearance of the *aurora* being so much more frequent now than formerly in these parts; if the earth's magnetic poles be like the centres of the aurora, as the phenomena indicate, it is plain the aurora must move along with them, and appear or disappear at places, according as the magnetic poles approach or recede from them; and hence it may be presumed that the earth's magnetic pole in the northern hemisphere is nearer the west of *Europe* in this century than it was in the last or preceding" (ibid. pp. 173-174).

<p align="center">7. Annual variation of aurorae</p>

Dalton considered the possibility of lunar effects on aurorae, stating "…it occurred to him that the phenomena had more frequently happened about the change of the moon than at any other time; this produced the suspicion that the aerial tides occasioned by the moon might have some influence upon it." (ibid. p. 175). When comparing the spring and neap tides, he noted annual variation, including maxima in the Kendal auroral data in April, October, and November, and minima in July and December (ibid. p. 242). In the 1834 edition, he proposed an alternative cause, stating that the aurorae occurred most frequently "in those months when the transitions of temperature are most rapid" (ibid. p. 242). In any event, both explanations are likely to be connected to the orbital/solar relationships utilized in contemporaneous theories.

## 8. Dalton's summary

Dalton summarized his findings as follows: "

The aurora consists of luminous arches or rings, drawn round the magnetic poles, in the manner of parallels of latitude around the poles of the earth on a terrestrial globe, and of luminous beams arising from them or amongst them, nearly perpendicular to the surface of the earth, or rather parallel to the dipping needle at the subja cent places. These concentric arches extend to 20°, sometimes 30°, but very rarely to 40° from the magnetic poles: thus, the aurora is seldom seen to the south, in Iceland, which is about 25° from the magnetic north pole; still more rarely in the Orkney and Shetland islands, which are 35° from the pole; and very rarely over the middle of Great Britain and Ireland, which is nearly 40° from the pole: thus, in the list recently given of one hun dred and eighty - four auroræ, only five or six arches were seen to pass the zenith , in this country . Au roræ are more numerous in the State of New York than in Britain, because that State is only 30° from the magnetic pole" (ibid. pp. 240-241).

## 9. Modern Interpretations

Dalton's implications are still valid and tell us volumes regarding the solar activity during the Dalton Minimum. Figure 3 presents the secular variation in the geomagnetic latitudes (MLATs) of Kendal and Manchester, where Dalton conducted his observations based on the GUFM1 model (Jackson *et al*., 2000). Their MLATs were approximately 2° closer to the geomagnetic pole than those in 1990 and have progressed steadily toward the equator since then. Therefore, observers at Kendal and Manchester had a greater opportunity to see auroral displays in the past, as long as the solar activity remained at the same magnitude. However, as shown in Figure 2, the auroral frequency decreased significantly after the end of Solar Cycle 4 (April 1798; Table 2 in Hathaway, 2015) and remained extremely low until 1826. The reported auroral frequencies still exhibit cyclic activity with apparent peaks at 1804 and 1814, slightly before the existing maxima of Solar Cycle 5 (February, 1805) and Solar Cycle 6 (May, 1816) in Table 1 in the paper by Hathaway (2015).

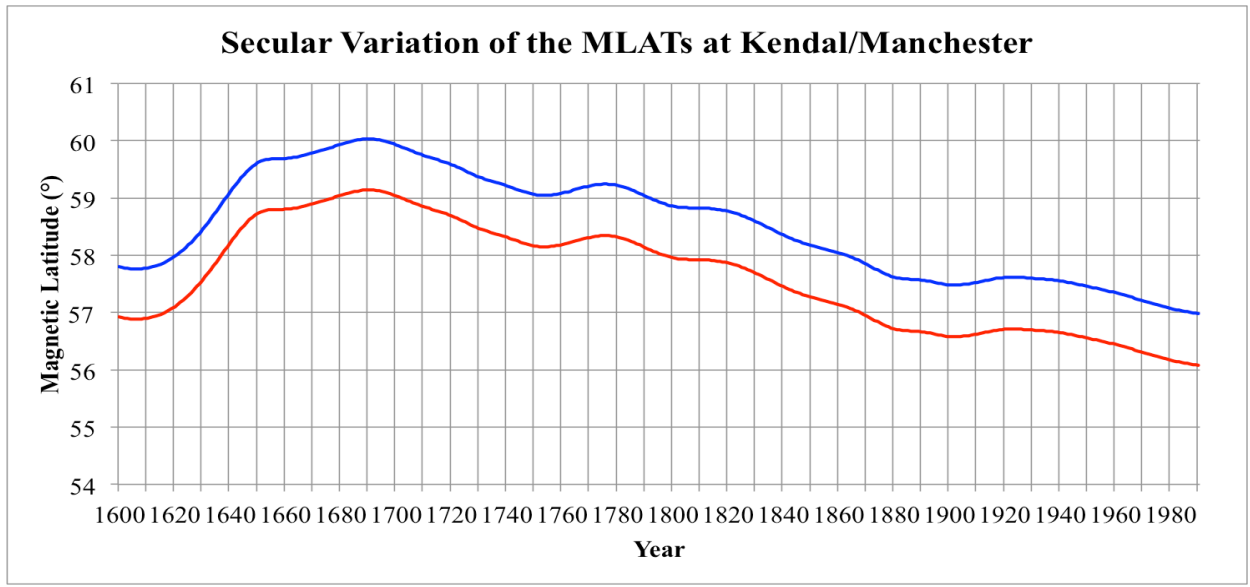

Figure 3. Secular variation of MLATs at Kendal (blue) and Manchester (red) from 1600 to 1990.

Around the Dalton Minimum, the reported auroral frequencies peaked in 1788 and 1830, which agrees with the existing maxima of Solar Cycle 4 (February, 1788) and Solar Cycle 7 (November, 1829) in Table 1 in the paper by Hathaway (2015). Regarding the cycle amplitude, Solar Cycle 4 exhibited aurorae much more frequently than Solar Cycle 7. Although Dalton moved equatorward from Kendal to Manchester in 1793, this contrast is still striking and confirms the greater amplitude of Solar Cycle 4, which agrees with existing reconstructions of sunspot numbers (Clette and Lefevre, 2016) and a proxy-based heliospheric magnetic field (McCracken and Beer, 2015). Additionally, Dalton's data show a local spike in 1797. This chronologically coincides with the lost cycle suggested by Usoskin *et al.* (2009), where sunspot positions may indicate another weak solar cycle at the end of Solar Cycle 4.

## 10. Conclusions

Dalton was a researcher of broad interest. Almost everything he observed required an explanation, whether it was botanical, meteorological, or chemical. His procedure was to first collect data. These were either his own observations, often using instruments of his own construction, or the published observations of others. These data were then classified and organized. Finally, he would search for a theoretical explanation for the observed data. As one might expect, his explanations relied on the knowledge available in his own time. Therefore, knowledge of the existence of the electron, which was discovered almost a century later, was not available to him. However, the questions he asked were still important.

Silverman and Hayakawa (2021) The Dalton Minimum and John Dalton's Auroral Observations, *Journal of Space Weather and Space Climate*. DOI: 10.1051/swsc/2020082

Dalton's life straddled the prolonged solar minimum at the beginning of the 19th century. The breadth of his geophysical interests and his groundbreaking systemization of atomic theory easily justify the naming of this solar minimum after him.

Therefore, the Dalton Minimum was named after Dalton following discussions by John Eddy, George Siscoe, and Sam Silverman. Since this informal discussion, significant scientific progress has been achieved over the past few decades. The "Dalton Minimum" has become accepted in the solar community and has been analyzed in comparison to the Maunder Minimum (*e.g.*, Clette *et al*., 2014; Usoskin, 2017). Analyses of contemporary records have shown sunspot numbers with distinct cyclicity, sunspot distributions in both solar hemispheres, and solar coronal streamers that are sufficiently bright for visual observations (Hayakawa *et al*., 2020a, 2020b). In contrast, the Maunder Minimum showed extreme suppression of solar cycles, southward concentration of sunspot distributions, and an apparent loss of coronal streamers (Eddy, 1976; Usoskin *et al*., 2015; Riley *et al*., 2015; Owens *et al*., 2017; Hayakawa *et al*., 2020c). However, there are a number of problems related to the current understanding of the Dalton Minimum, including the inconsistent reconstruction of solar cycles (*e.g.*, Svalgaard and Schatten, 2016; Muñoz-Jaramillo and Vaquero, 2019), even when accounting for one solar cycle lost in the early portion of the Dalton Minimum (Usoskin *et al*., 2009; Karoff *et al*., 2015; Owens *et al*., 2015). It is clear that these historical testimonies deserve further analysis. Because we have likely persisted through another small solar cycle, it may be worthwhile to reconsider Dalton's observations to understand how the sun behaves during a prolonged minimum.

## Acknowledgment

We thank Rachel Rosenblum for her cordial support on our research activity. We appreciates financial supports of JSPS Grant-in-Aids JP20K22367, JP20K20918, and JP20H05643, JSPS Overseas Challenge Program for Young Researchers, the 2020 YLC collaborating research fund, and the research grants for Mission Research on Sustainable Humanosphere from Research Institute for Sustainable Humanosphere (RISH) of Kyoto University and Young Leader Cultivation (YLC) program of Nagoya University. This work has been partly merited from participation to the International Space Science Institute (ISSI, Bern, Switzerland) via the International Team 417 "Recalibration of the Sunspot Number Series" and ISWAT-COSPAR S1-01 team.

Silverman and Hayakawa (2021) The Dalton Minimum and John Dalton's Auroral Observations, *Journal of Space Weather and Space Climate*. DOI: 10.1051/swsc/2020082

Jackson, A., Jonkers, A. R. T., Walker, M. R. (2000) Four centuries of geomagnetic secular variation from historical records, *Roy Soc of London Phil Tr A*, **358**, 957. DOI: 10.1098/rsta.2000.0569

Karoff, C., Inceoglu, F., Knudsen, M. F., Olsen, J., Fogtmann-Schulz, A. (2015) The lost sunspot cycle: New support from $^{10}$Be measurements, Astronomy & Astrophysics, 575, A77. DOI: 10.1051/0004-6361/201424927

Krivsky, L., and K. Pejml, (1988) World list of polar aurorae <55° and their secular variations, Part II, Astron. Inst. Czech. Acad. Sci., 75, 32-68.

Landscheidt, T. (1981) Swinging sun, 79-year cycle, and climatic change, *Journal of Interdisciplinary Cycle Research*, **12**, 3-19, DOI: 10.1080/09291018109359720

Lockwood, M., Barnard, L. (2015) An arch in the UK Mike Lockwood, Luke Barnard *Astronomy & Geophysics*, **56**, 4.25–4.30, DOI: 10.1093/astrogeo/atv132

Lovering, J., (1867) On the secular periodicity of the Aurora Borealis, Mem. Am. Acad. Arts Sci., 9, 101-111.

Maunder, E. W. (1922) The Prolonged Sunspot Minimum, 1645-1715, *Journal of the British Astronomical Association*, **32**, 140-145.

McCracken, K. G., Beer, J. (2015) The Annual Cosmic-Radiation Intensities 1391 - 2014; The Annual Heliospheric Magnetic Field Strengths 1391 - 1983, and Identification of Solar Cosmic-Ray Events in the Cosmogenic Record 1800 – 1983, *Solar Physics*, **290**, 3051-3069. DOI: 10.1007/s11207-015-0777-x

Millington, J.P., (1906) John Dalton, London: J.M. Dent & Co.

Miyake, F., Usoskin, I. G., Poluianov, S. (2019) Extreme Solar Particle Storms; The hostile Sun, Bristol, UK, IOP Publishing. DOI: 10.1088/2514-3433/ab404a

Muñoz-Jaramillo, A., Vaquero, J. M. (2019) Visualization of the challenges and limitations of the long-term sunspot number record, *Nature Astronomy*, **3**, 205-211. DOI: 10.1038/s41550-018-0638-2

Owens, M. J., Lockwood, M., Riley, P. (2017) Global solar wind variations over the last four centuries, *Scientific Reports*, **7**, 41548. DOI: 10.1038/srep41548

Owens, M. J., McCracken, K. G., Lockwood, M., Barnard, L. (2015) The heliospheric Hale cycle over the last 300 years and its implications for a "lost" late 18th century solar cycle, *Journal of Space Weather and Space Climate*, **5**, A30. DOI: 10.1051/swsc/2015032

Riley, P., Lionello, R., Linker, J. A., *et al.* (2015) Inferring the Structure of the Solar Corona and Inner Heliosphere During the Maunder Minimum Using Global Thermodynamic
17